\documentstyle[11pt,newpasp,twoside,epsf]{article}
\def\edcomment#1{\iffalse\marginpar{\raggedright\sl#1\/}\else\relax\fi}
\reversemarginpar

\begin{document}
\title{How common are Earths? How common are Jupiters?}
\author{Charles H. Lineweaver, Daniel Grether \& Marton Hidas}
\affil{School of Physics, University of New South Wales and the\\
 Australian Centre for Astrobiolgy, Sydney, Australia\\
charley@bat.phys.unsw.edu.au}
\begin{abstract}
Among the billions of planetary systems that fill the Universe, we would like to 
know how ours fits in. Exoplanet data can already be used to address the 
question: How common are Jupiters?  Here we discuss a simple 
analysis of recent exoplanet data indicating that Jupiter is a typical 
massive planet rather than an outlier. A more difficult question to address 
is: How common are Earths? However, much indirect evidence suggests that wet 
rocky planets are common.
\end{abstract}
\section{How Common are Jupiters? The statistics of massive detectable exoplanets}
Long before we detect Earth-like planets we will have a good general picture of the variety 
of massive planets in planetary systems.
Since Jupiter is the most prominent feature of our planetary system, and our
knowledge of other planetary systems is still rudimentary, we may, with current 
data reasonably hope to answer the less ambitious question:  How typical is 
Jupiter? 
The relevant analysis is now possible because a statistically 
significant sample is starting to emerge from which we can determine meaningful distributions 
in mass and period.
To quantify these distributions as accurately as 
possible, we have identified a subsample of exoplanets that is minimally
affected by the selection effects of the Doppler detection method (Fig. 1).
Within this subsample, after a simple completeness correction, we quantify trends in 
mass and period that are less biased than trends based on the full sample of 
exoplanets. Straightforward extrapolations of these trends, into the area of parameter space 
occupied by Jupiter, indicates that Jupiter lies in a region densely 
occupied by exoplanets (Lineweaver \& Grether 2002).  
Our analysis suggests that Jupiter is more typical than indicated by previous 
analyses. For example, instead of $M_{Jup}$ planets being twice as common as $2 M_{Jup}$ 
planets, we find they are 3 times as common.
The latest exoplanet data (detected between January and August
2002) supports and strengthens this conclusion (Lineweaver, Grether \& Hidas 2003).
Our claims for Jupiter 
being a typical massive planet are well-defined in terms of $Msin(i)$ and period but can not 
yet include orbital eccentricity since the eccentricity of most exoplanets is larger than the 
$\sim 0.1$ typical of our Solar System.

The frequency of Jupiter-like planets may have implications 
for the frequency of life in the Universe. 
A Jupiter-like planet shields 
inner planets from an otherwise much heavier bombardment by 
planetesimals, comets and asteroids 
\clearpage
\begin{figure}[h,t!]
\plotone{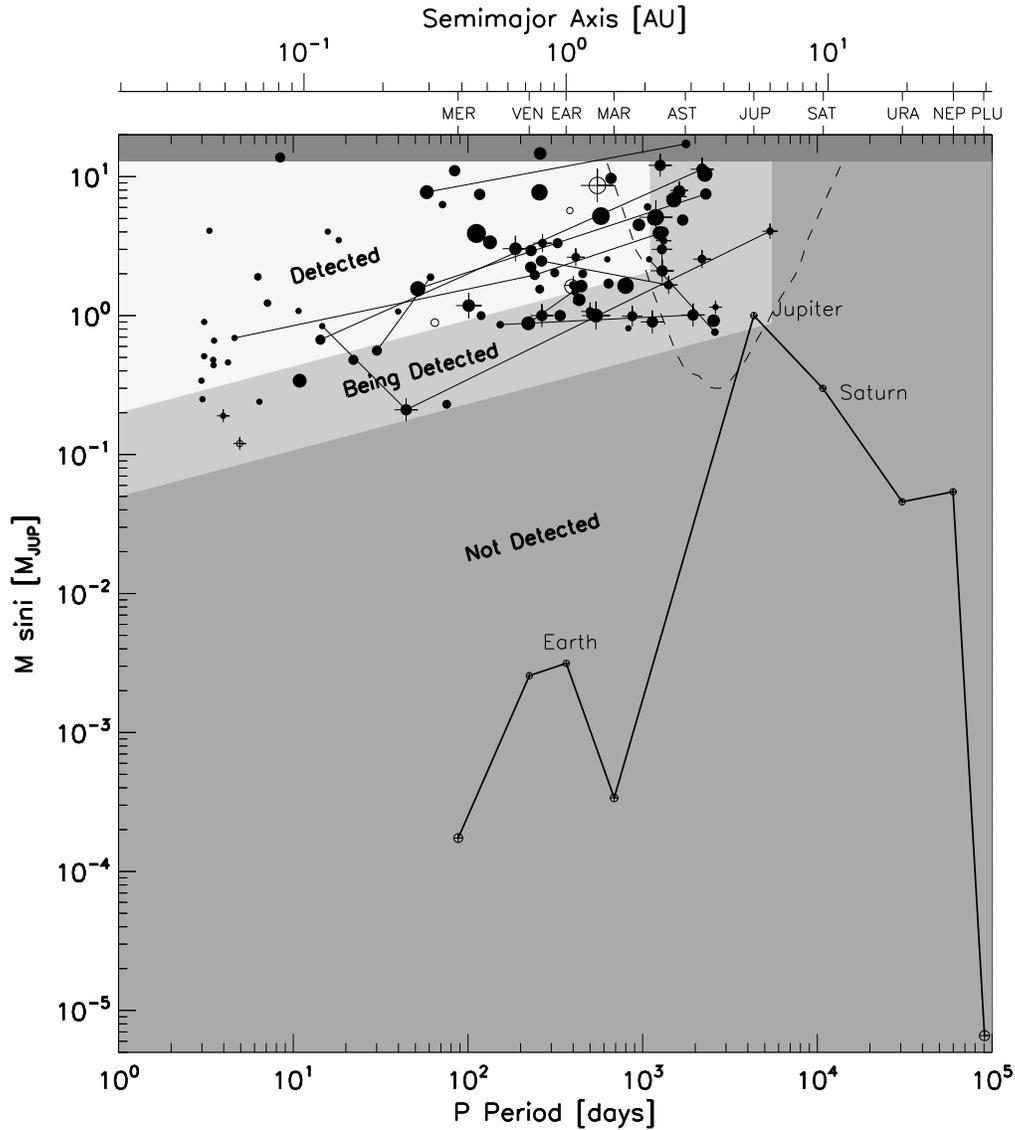}   
\caption{Comparison of the 101 exoplanets detected as of August 2002
to the mass and period of the planets of our Solar System.  
We would like to know how planetary systems in general are distributed 
in this plane. This figure shows that the Doppler technique has been able to sample a very specific high-mass, 
short-period region of the log P - log Msin (i) plane. Thus far, this sampled region does not overlap with 
the 10 times larger area of this plane occupied by the nine planets of our Solar System. Thus, there is room in 
the $\sim 95\%$ of target stars with no Doppler-detected planets, to harbor planetary systems like 
our Solar System. Null results from microlensing searches have been used to constrain the 
frequency of Jupiter-mass planets (Gaudi et al. 2002). Less than 33\% 
of the lensing objects (presumed to be Galactic bulge M-dwarfs) have planetary companions within the dashed 
wedge-shaped area (the period scale, but not the AU scale, is applicable to this area).
For more details see Lineweaver and Grether 2002. 
}
\end{figure}
\clearpage
\begin{figure}
\plotfiddle{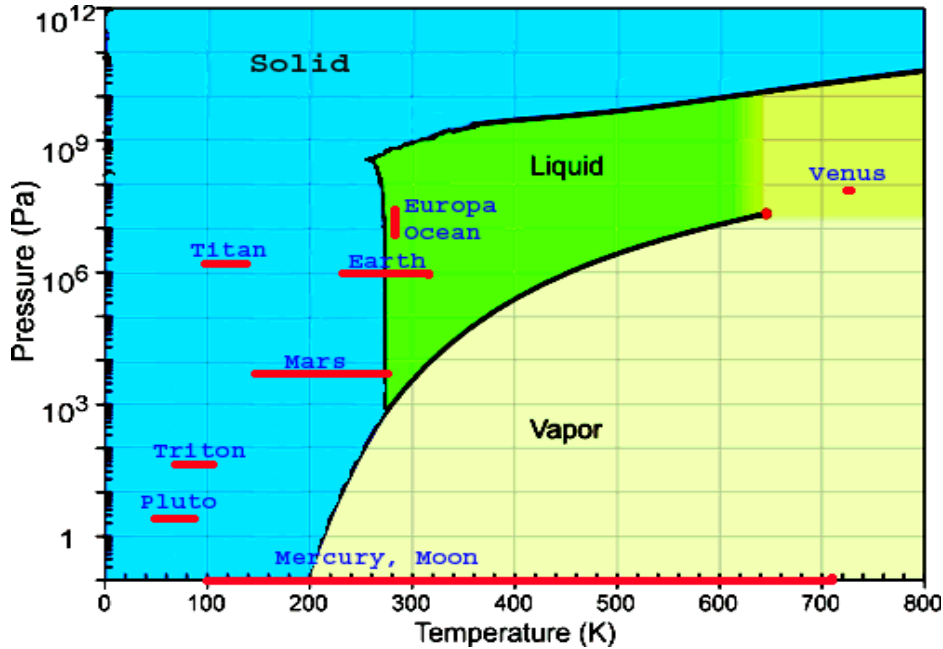}{8cm}{0}{70}{60}{-180}{0}
\caption{ How common is liquid water in the Universe? 
In this phase diagram of water the scatter of the bodies of our Solar System 
is large. In the absence of any systematic pattern
avoiding the liquid water region, and to the extent that other planetary systems share
such random scatter, we can expect wet rocky planets to exist in a substantial 
fraction of planetary systems in the Universe.
Diagram adapted from www.sbu.ac.uk/water/phase.html}
%
%
\end{figure}

\noindent during the first billion years
after formation of the central star.
Wetherill (1994) has estimated that Jupiter significantly reduced 
the frequency of sterilizing impacts on the early Earth during the important
epoch $\sim 4$ billion years ago when life originated on Earth.
The removal of comet Shoemaker-Levy by Jupiter in 1994, is a more recent 
example of Jupiter's protective role.
However, we know so little about the details of the biochemistry of biogenesis that
impacts may also play a constructive (not necessarily destructive) role in biogenesis.

\section{How Common are Earths?}
Indirect evidence for wet rocky planets being common includes
theoretical inferences from conservation of angular momentum in gravitationally collapsing objects and 
observations of the ubiquity of circumstellar disks and the appropriate-for-planet-formation time scales 
of their disappearance from around young stars (Hillenbrand et al. 2002).  
The high frequency of accreted (not-captured) moons in our Solar System also suggests that satellite 
formation is a run-of-the-mill product of star formation.
Rocky planets appear to be the most natural repositories of the remnant refractory material from 
high and average metallicity circumstellar disks.

Wetherill (1995) has suggested that Jupiters may be harder to form than 
rocky/terrestrials because a Jupiter may need a rocky core of $\sim 10 \; M_{Earth}$ in 
place very early in order to start runaway collapse of the gaseous (H, He) protoplanetary 
disk before it dissipates. Apparently it is not easy to develop a rocky core quickly enough.  
Since we find that Jupiters are probably typical, this implies that rocky/terrestrial planets are 
even more typical.

Water is probably the most common triatomic molecule in the Universe. The billions of dirty snowballs
orbitting the Sun are the remnants of a larger population which, for a billion years, heavily 
bombarded the inner Solar System, repeatedly adding a wet veneer to the rocky planets.
Such cometary bombardments are probably a universal feature of planet formation. 
Whether these veneers typically last long enough to permit biogenesis depends on the mass, 
atmosphere and orbital distance of the particular rock. 
Fig. 2. shows the phase diagram of water and the scatter of the phases of $H_{2}O$ gravitationally 
attached to some of the rocks of our Solar System.

The most plausible arguments mustered against the idea that Earths are common include the following:
planets  may not form if erosion, rather than growth, occurs during collisions of planetesimals 
(Kortenkamp \& Wetherill 2000). The present day asteroid belt may be an example of such non-growth. In 
addition, not all circumstellar disks produce an extant planetary system. Some fraction may spawn a 
transitory system only to be accreted by the central star along with the disk (Ward 1997). Also, observations 
of star-forming regions indicate that massive stars disrupt the protoplanetary disks around neighboring lower 
mass stars, aborting their efforts to produce planets (Henney \& O'Dell 1999).

We have been assuming that Earth-like planets are wet rocks. 
If we define Earth-like to mean only planets exactly like Earth, then of course the more detailed 
our description the less common such planets will be. 
The difficulty of differentiating details that are important for biogenesis from details that merely 
modify the evolutionary path of life undermine the relevance of including precise details in our 
description of ``Earth-like'' planet.

\end{document}